\documentclass[traditabstract]{aa} % for the abstract without structuration 
\usepackage{graphicx}
\usepackage{txfonts}
\usepackage{natbib}

\def\Mpc{Mpc}
\def\hMpcm{\,h\Mpc^{-1}}

\def\kperp{k_\perp}
\def\kpar{k_\parallel}
\def\koperp{k_{BAO\perp }}
\def\kopar{k_{BAO\parallel}}

\begin{document}
%
%\title{ Reconstruction of HI power spectra}
\title{ Reconstruction of HI power spectra with radio-interferometers to study dark energy}

   \author{R. Ansari\inst{1}, J.-M. Le Goff\inst{2}, Ch. Magneville\inst{2}, 
M. Moniez\inst{1}, N. Palanque-Delabrouille\inst{2}, J. Rich\inst{2}, \\ V. Ruhlmann-Kleider\inst{2} 
          \and Ch. Y\`eche\inst{2}
         	          }

   \institute{Laboratoire de l'Acc\'el\'erateur Lin\'eaire, IN2P3-CNRS, Universit\'e de Paris-Sud, BP. 34, 91898 Orsay Cedex, France\\
                       \and
           CEA, Centre de Saclay, IRFU,  F-91191 Gif-sur-Yvette, France 
             }

   \date{Received June 15, 2008; accepted yyyy xx, 2008}

% \abstract{blbl}{clcl}{dldl}{elel}{flfl} 
% 5 {} token are mandatory

\authorrunning{R. Ansari et al.}
\titlerunning{ Reconstruction of HI power spectra with radio-interferometers to study dark energy}
 
  \abstract{
Among the tools available for the study of the dark energy driving the
expansion of the Universe, Baryon Acoustic Oscillations (BAO) 
and their effects on the matter power spectrum
are particularly attractive.
It was recently proposed to study these oscillations
by mapping the  21cm emission 
of the neutral hydrogen in the redshift range $0.5<z<3$.
We discuss here the precision of such measurements 
using radio-interferometers consisting of arrays 
of dishes or north-south
oriented cylinders. 
We then discuss the resulting uncertainties on the BAO scales and
the sensitivity to the parameters of the 
Dark Energy equation of state. 
}
  

   \maketitle
%
%________________________________________________________________

%\input{intro}

\section{Introduction}

One of the goals of current cosmological research is to
fully characterize the ``dark energy'' that drives the apparent acceleration
of the expansion of the Universe.
Among the available tools \citep{DETF}, much interest has been
generated by the features in the matter power spectrum that result
from Baryon Acoustic Oscillations (BAO) in the pre-recombination
Universe.  These oscillations cause a peak in the matter correlation function
at a comoving distance, $s\sim105h^{-1}Mpc$, equal to the acoustic horizon at
recombination. 
Equivalently, the BAO induce ``wiggles'' in the matter power spectrum
with peaks
at comoving wave numbers $k=(n+1/2)\pi/s$ for $n=2,4,....$.
The theoretical  positions of these peaks  are known 
to a precision of order 1\% so they
can be used as  reliable standard rulers
to study the expansion history.

After first being seen in the CMB anisotropy spectrum
\citep{boomerang,wmap},
BAO effects were subsequently observed
in the low-redshift ($z<1$) 
galaxy correlation
function and power spectrum by the SDSS survey \citep{BAO2,SDSSPk} 
and the 2dGFRS survey \citep{BAO1}. 
With future data in the redshift range $0.5<z<3$, 
it is hoped to provide precise 
constraints on dark-energy
parameters. 
It has been proposed to do this with redshift
surveys using galaxy optical spectra \citep{wfmos}
or HI 
%(21cm, $\nu_0=1.42GHz$) 
emission \citep{SKA}
or by mapping in three dimensions the pattern of Lyman-$\alpha$ absorption
of distant quasars \citep{lyalpha}.

An alternative elegant approach was recently proposed by 
\citet{HSHSMoriond} and further developed by \citet{HSHSPRL}.
They propose to use neutral hydrogen (HI) as a tracer of matter and
simply map out the 21cm emission with an angular resolution
that is insufficient to detect individual galaxies (where most HI 
is concentrated) but sufficient
to observe the BAO wiggles.
At redshift $z=1.5$, the acoustic horizon subtends 0.0334rad on the sky
which can be resolved with telescopes of size of order 100m.  
These would be considerably smaller and cheaper than the 1km elements
necessary to detect individual high redshift galaxies \citep{SKA}.

Mapping the matter distribution using HI 21 cm emission as a tracer has
been extensively discussed in literature (e.g. \citet{Furlanetto,fftt}). 
Several projects, such as LOFAR \citep{Rottgering}
or MWA \citep{Bowman} aim at detecting the reionization epic ($z\sim10$). 
Detecting BAO features around
$z\sim1$ using HI radio emission has also been discussed by \citet{Wyithe}.

In Section \ref{pssec} we review the expectations for the HI power
spectrum $P_{HI}(k)$ and its relation to the large-scale structure
matter power spectrum $P_{LSS}(k)$.  
Section \ref{datasec} describes the interferometric observations that
are assumed for this study and 
Section \ref{recsec} presents a simplified procedure
to use these observations to 
reconstruct the HI power spectrum in a small volume
near the zenith.
While this procedure is not entirely realistic, it
allows us to give estimates of the 
reconstructed noise power spectrum, $P_{noi}(k)$
(Fig. \ref {powerfig}).
Section \ref{radsourcesec} discusses the problem of subtraction of
foreground and background radio sources.
Finally, in Section \ref{cosmosec} we describe the constraints on the
cosmological parameters that can be derived from the determination
of the BAO peaks in the power spectrum.

\section{The HI power spectrum}
\label{pssec}

The mean HI brightness temperature (assumed to be much greater
than the CMB temperature) is \citep{h1temp}
\begin{equation}
\bar{T}_{HI}(z) = 
%\frac{\lambda^2 I}{2k_B} = 
0.0466mK
\frac{(1+z)^2f_{HI}(z)}{[\Omega_\Lambda+(1+z)^3\Omega_M]^{1/2}}\,
\frac{\Omega_{HI}h_{70}}{3.5\times10^{-4}}
\label{meantemp}
\end{equation}
where $\Omega_{HI}\sim 3.5\times10^{-4}$ \citep{HIPASS}
is the current mean cosmological HI density 
and
\begin{equation}
f_{HI}(z)=\frac{(HI/H)_z}{(HI/H)_{z=0}}
\end{equation}
is the fraction of hydrogen in atomic form relative to this fraction
at $z=0$.
Studies of Lyman-$\alpha$ absorption \citep{lyalpha}
 indicate the $f_{HI}(z=1.5)\sim 3$ implying
$\bar{T}_{HI}(z=1.5)=0.00033K$.
This is three orders of magnitude  below the brightness of
extragalactic radio source emission, 
$\sim 3400Jy\,sr^{-1}$,
corresponding to 
$\bar{T}_{rs}\sim0.3K$. 
However, these sources, as well as Milky Way emission and CMB radiation,
have smooth frequency distributions, allowing them, in principle, to 
be subtracted.

We are interested in the spatial variations of the HI temperature
about the mean.
Consider a small cube of volume $V$ at mean redshift $z$ corresponding
to a HI frequency $\nu=\nu_0/(1+z)$.
The cube covers  
a   solid angle $\Delta\Omega=\Delta\alpha\times\Delta\delta$
($\alpha=$ r.a., $\delta=$ dec.$\sim0$)
and redshift range $\Delta z $ corresponding to 
a frequency range  
$\Delta \nu=\nu_0\Delta z/(1+z)^2$. 
The present volume of the cube is
\begin{equation}
V=d_T^2d_H\Delta\Omega\,\Delta z
=\frac{(c/H_0)d_T^2 \Delta\Omega \,\Delta z}
{\sqrt{\Omega_\Lambda+\Omega_M(1+z)^3}}
%\frac{\Delta\Omega \Delta\nu}{\nu_0}
\label{cubevolume}
\end{equation}
where for a flat $\Lambda$CDM universe, the comoving angular distance $d_T$
and Hubble distance $d_H$ are
\begin{displaymath}
d_T = \int_{0}^z\frac{(c/H_0)\,dz}{
\sqrt{\Omega_\Lambda+\Omega_M(1+z)^3}}
\hspace*{5mm}
d_H = \frac{c/H_0}
{\sqrt{\Omega_\Lambda+\Omega_M(1+z)^3}} \;.
\label{dTdH}
\end{displaymath}
%($d_T$ is the geometric mean of the luminosity and angular-size distance and
%$d_H$ is the Hubble distance.)

Inside the cube, the 
sky brightness can be expanded using functions that satisfy
periodic boundary conditions.  
For small cubes ($\Delta\alpha,\, \Delta\delta, \,\Delta z \ll 1$) the
functions can be taken to be complex exponentials of $(\alpha,\delta, z)$: 
\begin{equation}
T_{HI}(\alpha,\delta,z ) \;=\; 
\frac{\bar{T}_{HI}}{\sqrt{V}} \sum_{\vec{n}}\Gamma_{HI}(\vec{n})\, e^{2\pi i
(
n_\alpha \alpha/\Delta\alpha +
n_\delta \delta/\Delta\delta +
n_z z/\Delta z 
)}
\label{skytemp}
\end{equation}
\nonumber
where the $\Gamma_{HI}(\vec{n}=(n_\alpha,n_\delta,n_z))$, 
are defined for integer $n_\alpha,n_\delta,n_z$. 
The wave vector $\vec{k}$ is related to $\vec{n}$ by
\begin{equation}
k_\alpha=\frac{2\pi n_\alpha}{d_T \Delta\alpha} \hspace*{10mm}
k_\delta=\frac{2\pi n_\delta}{d_T \Delta\delta} \hspace*{10mm}
k_z=     \frac{2\pi n_z   }{d_H \Delta z}
\end{equation}
The HI power spectrum is:
\begin{equation}
P_{HI}(k) \;=\; \langle |\Gamma_{HI}(\vec{k})|^2\rangle
\end{equation}
the average being over the 
modes $\vec{k}$ with wave numbers near $k$.
The number of such modes in the cube and in the interval
$\Delta k$ is:
%\begin{eqnarray}
\begin{displaymath}
%\label{nmodes}
N_{\Delta k} =  \frac{V k^3}{4\pi^2}\,\frac{\Delta k}{k}
= \frac{(c/H_0)d_T(z)^2\Delta\Omega\Delta z k^3}
{4\pi^2\left[\Omega_\Lambda + (1+z)^3\Omega_M   \right]^{1/2}}
\,\frac{\Delta k}{k}
%\hspace*{20mm}
\nonumber
\end{displaymath}
\begin{eqnarray}
\label{nmodes}
\hspace*{10mm}
=7.25\times10^4
\left(\frac{d_T(z)}{c/H_0}\right)^2
\frac{\Delta\Omega \Delta z}{2\pi/5}
\left(\frac{k}{0.075h\,Mpc^{-1}}\right)^3
\frac{\Delta k/k}{0.2}
\nonumber
\\
 \times \left[\Omega_\Lambda + (1+z)^3\Omega_M   \right]^{-1/2}
\hspace*{20mm}
\end{eqnarray}
Since the $|\Gamma(k)|^2$ are random number with variance equal to $P(k)$ 
the precision with which $P(k)$ can be measured is $P(k)/\sqrt{N_{\Delta k}}$.

The HI power spectrum is expected to be similar to the galaxy
power spectrum.
Figure \ref{powerfig} shows the expected galaxy 
spectrum normalized to agree with that measured by  SDSS
at $z\sim0.1$ \citep{sdssps}.
Also shown is 
this spectrum extrapolated to $z=1.5$ assuming 
$\Omega_M,\Omega_\Lambda=0.27,0.73$.  
The spectrum has BAO peaks at
\begin{equation}
k_n =(n+1/2)\pi/s = 0.075 hMpc^{-1} \frac{n+1/2}{2.5}
\hspace*{5mm} n=2,4,6.....
\end{equation}
where $s=105h^{-1}Mpc$ is the sound horizon at recombination.
The relative crest to trough amplitudes at the first three peaks 
are expected to be
$\Delta P/P\sim0.13$,
$\Delta P/P\sim0.09$, and 
$\Delta P/P\sim0.05$.

\section{Interferometric Observations}
\label{datasec}

\begin{figure}[ht]
\includegraphics[width=8.5cm]{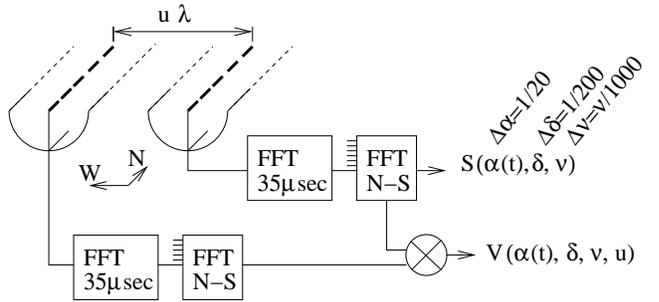}
\caption{ 
Two cylinders instrumented with dipole receiving antennae
along their axes.
After a sampling time $t_{int}$ (taken here to be $35\mu sec$),
an FFT is performed for each receiver to separate frequency, $\nu$,
components.
Receivers within a given cylinder are then combined by (north-south) FFT
to form beams in declination, $\delta$.
The amplitudes from the two cylinders can then be combined
to form visibilities, $V(\alpha(t),\delta,\nu,{\bf u})$ where
$\alpha(t)$ is the right-ascension determined by the time of the
observation and where
the cylinder separation is ${\bf u}\lambda$.
}
\label{reconfig}
\end{figure}

Observations of GHz radiation can be performed with 
interferometers consisting of arrays
of reflectors and receivers.
In this paper, we consider two types of arrays. The first, 
drawn schematically in Fig. \ref{reconfig},
consist of   north-south
oriented cylindrical reflectors pointing towards the zenith.  
The reflectors have dipole receiving antennae deployed along
their axes. 
The information
to be extracted from a pair of such cylinders is illustrated
in Fig. \ref{reconfig}.
The amplitude for each receiver is sampled over a time 
$t_{int}$ 
after which a Fast-Fourier Transform (FFT) is performed 
to separate frequency components.
We take $t_{int}$ to be
sufficiently long ($\sim35\mu sec$) to give excellent resolution
on the BAO features in the radial direction.
The receivers within a given cylinder are then combined by a FFT
to form beams covering  pixels
in declination of width $\sim\lambda/D_\delta$ for a wavelength $\lambda$
and cylinder length $D_\delta$. 
This requires that receivers be spaced by $\lambda/2$  
in order to avoid large side-lobe contamination.
The number
of receivers per cylinder is thus of order 500 for $D_\delta=100m$.
The amplitudes from two cylinders separated
in ${\bf u}\lambda$ in the east-west direction can than be correlated
to form a ``visibility'', $V(\alpha(t), \delta, \nu, {\bf u})$ that
is defined for discrete values of declination, $\delta$, and
frequency, $\nu$, defined by the two FFT's.  The right ascension,
$\alpha(t)$ is defined by the time of observation, $t$.

\begin{figure}[ht]
\includegraphics[width=8.5cm]{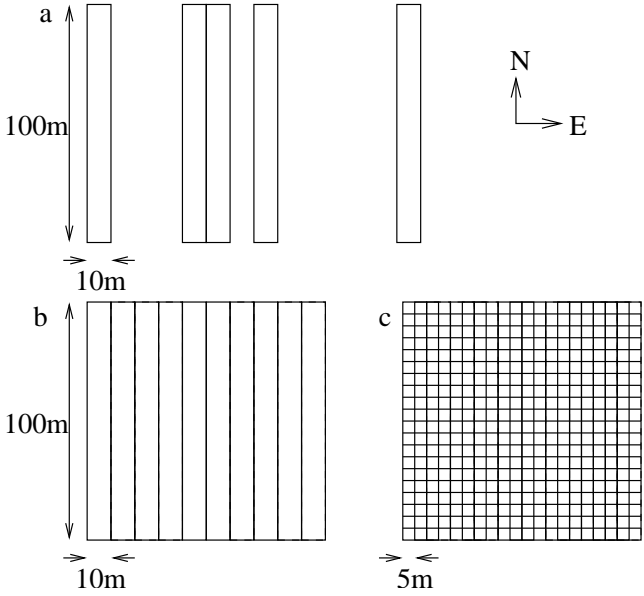}
\caption{ 
The three telescope configurations considered:  
a: ``unpacked cylinders'';
b: ``packed cylinders'';
c: ``packed dishes''.
In the unpacked array, the five cylinders are separated 
from the most western cylinder by
distances of $n\times 10m$ with $n=0,4,5,7,13$.
}
\label{configfig}
\end{figure}

\begin{figure}[ht]
\includegraphics[width=8.5cm]{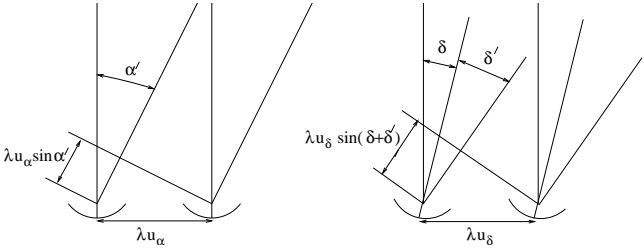}
\caption{ 
The pointing configuration of telescope pairs (at the terrestrial
equator for simplicity).  
The left
panel shows two cylinders separated in the east-west direction
and pointing toward the zenith.
The right
panel shows two dishes separated in the north-south direction
and pointing in a direction $\delta$ from the zenith.
}
\label{pointfig}
\end{figure}

The second type of array considered in this paper consists of dishes
the are orientable in declination.  The readout for pairs of
dishes is the same as for the cylinders in Figure \ref{reconfig}
except that there is no north-south FFT so that the declination
$\delta$ is defined by the common dish pointing.  Additionally,
the dish separation has components in the north-south and east-west
directions so ${\bf u}$ is a vector.

The cylinders of Fig. \ref{reconfig} give high resolution information
in the declination and frequency  directions.
In order to provide high quality information in the right-ascension
direction, it is necessary to have several cylinders with 
a range of reflector separations $\lambda{\bf u}$.
Schematics of 
the three arrays considered here are shown in Figure \ref{configfig}.
Two arrays, $a$ and $b$, are arrays of fixed cylindrical  reflectors
of width $D_\alpha =10m$ and length $D_\delta=100m$.
The east-west configuration of the five cylinders in configuration
$a$ is chosen to give uniform sensitivity over the 
$k$ range necessary for BAO studies.  
As such, they are placed at positions that are
integral multiples of the cylinder width as shown in the figure.
On the other hand, the ten cylinder in configuration $b$
are adjacent to form a ``packed array'' giving redundant information
over the interesting $k$ range.

The third system, c, is an array of $D_\alpha\times D_\delta=5m\times5m$
dishes (taken to be squares for mathematical convenience).
Each dish is instrumented with one receiver.
The field-of-view in both angular directions is
$\sim\lambda/D_\alpha$.
In order to survey the sky, 
the dishes must be  pointable in declination. 
As for the cylinder arrays, 
the right-ascension is determined by the time of observation.

As  illustrated in Fig. \ref{reconfig},
each readout-FFT sequence generates a set of amplitudes $S_{adfc}$
indexed by the four integers:  
$a$ giving the right ascension, $\alpha_a$,
determined by the time of observation; 
$(d,f)$ denoting the set of amplitudes
generated by FFT corresponding to pixels in declination-frequency
space centered on $\delta_d,\nu_f$ (only one $d$ per readout 
for dishes); 
and $c$ defining the reflector (dish or cylinder).
The amplitudes are the sum of signal and noise:
\begin{equation}
S_{adfc} \;=\; S^{noi}_{adfc} +S^{sig}_{adfc} 
%\;=\; S^{noi}_{adfc} +S^{HI}_{adfc} +S^{rs}_{adfc} 
\end{equation}

The angular and frequency response are determined 
by the lobe functions, 
$L_{adfc}(\alpha^\prime,\delta^\prime ,\nu^\prime)$,
peaked at $\alpha^\prime=\delta^\prime=\nu^\prime=0$
and normalized
so that the expectation value of the  squared signal is
\begin{displaymath}
\langle |S^{sig}_{adfc}|^2 \rangle = 
\int d\alpha^\prime d\delta^\prime  d\nu^\prime  
L(\alpha^\prime,\delta^\prime,\nu^\prime) 
%L_{adfc}(\alpha^\prime,\delta^\prime,\nu^\prime) \,
T(\alpha^\prime+\alpha_a,\delta^\prime+\delta_d,\nu^\prime+\nu_f)
\end{displaymath}
For the rest of this note we make the approximation
that $L_{adfc}$ is independent of $(a,d,f,c)$.
The noise contribution is normalized so that the mean 
signal-to-noise
is given by $\bar{T}/T_{sys}$
\begin{equation}
\langle |S^{noi}_{adfc}|^2\rangle \;= 
T_{sys} \int d\alpha^\prime\, d\delta^\prime\, d\nu^\prime\,
 L(\alpha^\prime,\delta^\prime,\nu^\prime) 
\end{equation}

The visibility for direction $(a,d)$ and frequency
$f$ for cylinders (or dishes) $c$ and $c^\prime$ separated in space by 
$\vec{u}\lambda_f$ is:
\begin{equation}
V_{adf\vec{u}} = S_{adfc}S^*_{adfc^\prime} 
\end{equation}
The noise-noise contribution is a random number of
vanishing mean and variance
determined by $T_{sys}$:
%\begin{equation}
%\left\langle V^{noi}_{adf\vec{u}}\right\rangle =0 
%\end{equation}
\begin{equation}
%\hspace*{10mm}
\left\langle| V^{noi}_{adf\vec{u}} |^2\right\rangle^{1/2} =
T_{sys} \int d\alpha^\prime\, d\delta^\prime\, d\nu^\prime\,
L(\alpha^\prime,\delta^\prime,\nu^\prime) 
\label{noisevisibility}
\end{equation}
Following Fig. \ref{pointfig},
the  signal-signal contribution to the visibility 
has an expectation value
\begin{eqnarray}
\langle V_{adf\vec{u}}^{sig} \rangle = &
\int d\alpha^\prime d\delta^\prime d\nu^\prime
L(\alpha^\prime,\delta^\prime,\nu^\prime)\,
T(\alpha^\prime+\alpha_a,\delta^\prime+\delta_d,\nu^\prime+\nu_f)
\nonumber
\\
&
\times
\exp \left[-2\pi i(u_\alpha \sin\alpha^\prime + u_\delta\sin(\delta_d+\delta^\prime))\right]
\end{eqnarray}
For $\alpha^\prime,\delta^\prime\ll 1$ and 
$\delta_d\ll 1$  this simplifies to
\begin{eqnarray}
\label{vissmallangle}
\langle V_{adf\vec{u}}^{sig} \rangle e^{2\pi iu_\delta \delta_d} \;=\; 
\hspace*{50mm}
\\ 
\int d\alpha^\prime\,d\delta^\prime\,d\nu^\prime\, 
L(\alpha^\prime,\delta^\prime,\nu^\prime)\,
T(\alpha^\prime+\alpha_a,\delta^\prime+\delta_d,\nu^\prime+\nu_f)
e^{-2\pi i\vec{u}\cdot \vec{\theta}^\prime} \nonumber
\end{eqnarray}
where ($\vec{\theta}^\prime=(\alpha^\prime,\delta^\prime)$).

\section{Reconstruction of HI Power spectrum}
\label{recsec}

The visibilities, $V_{adf {\bf u}}$ can be  combined to give
information on the HI distribution.
For a given readout, a Fourier transform of $V_{adf {\bf u}}$
over ${\bf u}$ yields
a map of the field of view.  
We are more interested in the power spectrum which is
related to the Fourier transform of $V_{adf {\bf u}}$
over right-ascension, declination and frequency:
\begin{equation}
\tilde{V}(\vec{k},\vec{u})\;\equiv\;
\frac{1}{N_d N_f N_a}
\sum_{adf}e^{- i(
d_T k_\alpha \alpha_a +
d_T k_\delta \delta_d +
d_H k_z    z_f)}
V_{adf\vec{u}}e^{2\pi i u_\delta \delta_d} 
\label{vtildedef}
\end{equation}
where there are $N_d\times N_f\times N_a$ pixels densely and
uniformly covering the cube ($\Delta\alpha, \Delta\delta, \Delta\nu$).
In the absence of noise, substituting 
the expansion (\ref{skytemp})
into (\ref{vissmallangle}) shows that $\tilde{V}(\vec{k},\vec{u})$
is proportional to the product of 
$\Gamma_{HI}(\vec{k})$
and the Fourier transform of the lobe function.
More generally, the visibilities are due to noise and signal so we have
\begin{equation}
\tilde{V}(\vec{k},\vec{u})
\;=\; \frac{\bar{T}\Gamma(\vec{k})\, F(\vec{k},\vec{u})}{\sqrt{V}}
\int d\alpha^\prime\, d\delta^\prime\, d\nu^\prime  
L(\alpha^\prime,\delta^\prime,\nu^\prime )
%\hspace*{20mm} 
%\Gamma(\vec{k})=\Gamma_{HI}+\Gamma_{noi}+\Gamma_{rs} 
\label{vtildeformula}
\end{equation}
where $\Gamma(\vec{k})=\Gamma_{HI}(\vec{k})+\Gamma_{noi}(\vec{k})$ 
(plus foreground/background contributions)
and where
the dimensionless ``form factor'' for the mode $\vec{k}$ is
\begin{equation}
F(\vec{k},\vec{u})= \frac{
\int d\alpha^\prime\, d\delta^\prime\, d\nu^\prime  
L(\alpha^\prime,\delta^\prime,\nu^\prime ) 
%e^{ik_\delta d_T \delta^\prime }
e^{ik_z d_H (1+z)^2\nu^\prime } 
e^{i \vec{\theta}^\prime\cdot\left(d_T\vec{k} -2\pi \vec{u}\right)}
}{
\int d\alpha^\prime\, d\delta^\prime\, d\nu^\prime  
L(\alpha^\prime,\delta^\prime,\nu^\prime ) 
}\;.
\end{equation}
If $L$ is symmetric, $F$ is a real function.

The estimate of $\Gamma(\vec{k})$ 
based on  the visibility of one reflector pair with separation $\vec{u}$ 
is denoted $\Gamma(\vec{k},\vec{u})$.  From (\ref{vtildeformula}), it is
given by
\begin{equation}
\Gamma(\vec{k},\vec{u})\;=C
\frac{\tilde{V}(\vec{k},\vec{u})}{ F(\vec{k},\vec{u}) }
%\hspace*{20mm} 
\label{etaformula}
\end{equation}
where the ``calibration constant'', $C$, relates the 
power spectrum to the measurements
and form factors:
\begin{equation}
C\;= \frac{\sqrt{V}}
{ \bar{T} \int d\alpha^\prime\, d\delta^\prime\, d\nu^\prime  
L(\alpha^\prime,\delta^\prime,\nu^\prime )}
\end{equation}

Since more than one value of $\vec{u}$ can  be used to 
estimate $\Gamma(\vec{k})$,
one can make a weighted average of the $\Gamma(\vec{k},\vec{u})$ derived with
(\ref{etaformula}). 
In the next subsection, we will show that the noise power is inversely
proportional to $F(\vec{k},\vec{u})$ so it
is  reasonable to have weights that are increasing functions of
$|F(\vec{k},\vec{u})|\sim F(\vec{k},\vec{u})$.
It is most convenient to take the weights to be equal to $F(\vec{k},\vec{u})$:
\begin{equation}
\Gamma(\vec{k})\;=
\frac{ \sum_{\vec{u}} F(\vec{k},\vec{u}) \Gamma(\vec{k},\vec{u})}
{ \sum_{\vec{u}} F(\vec{k},\vec{u})}
\;=C 
\frac{\sum_{\vec{u}}  \tilde{V}(\vec{k},\vec{u})}
{\sum_{\vec{u}} F(\vec{k},\vec{u})}
\label{meanetaformula}
\end{equation}
where the sum is over those values of $\vec{u}$ with $F(\vec{k},\vec{u})\neq0$
(Other choices of weighting may give slightly lower noise power but at
the price on introducing wiggles in the noise power spectrum.)

\subsection{The noise}
\label{noisesec}

The noise contribution to the
$\tilde{V}(\vec{k},\vec{u})$  (\ref{vtildedef})
is the mean of  $N_aN_dN_f$ random numbers of variance
given by (\ref{noisevisibility}).
This gives 
$\left\langle \tilde{V}_{noi}(\vec{k},\vec{u})\right\rangle =0$
and
\begin{equation}
\left\langle| \tilde{V}_{noi}(\vec{k},\vec{u}) |^2\right\rangle^{1/2} =
\frac{
\left\langle| V_{adf\vec{u}}^{noi} |^2\right\rangle^{1/2}}
{\sqrt{N_aN_dN_f}} \hspace*{30mm}
\label{noisetildevisibility}
\end{equation}
The noise, 
$\Gamma_{noi}(\vec{k},\vec{u})$ reconstructed using (\ref{etaformula})
is inversely
proportional to $F(\vec{k},\vec{u})$ with an expectation value, $P_0$,
given by 
\begin{equation}
P_0\;\equiv\; 
\left\langle 
|\Gamma_{noi}(\vec{k},\vec{u})F(\vec{k},\vec{u})|^2 \right\rangle \;= 
\frac{T_{sys}^2}{\bar{T}^2}\,\frac{V}{N_d N_f N_a}  \; .
\label{pnoise}
\end{equation}
The survey volume to pixel ratio
$V/N_a N_d N_f$ depends on the interferometer configuration.
In all cases the number of pixels along
the frequency direction is  $N_f=t_{int}\Delta\nu$ where $t_{int}$ is
the integrating time between electronic readouts.
In the angular directions, cylinders and dishes differ.
Dishes have only one angular beam, so 
we have $N_a N_d=t_{tot}/t_{int}$ where $t_{tot}$
is the total observation time of the survey.
Cylinders, with FFT beam-forming,  have $N_{a}=t_{tot}/t_{int}$ and 
$N_d=\Delta\delta/(\lambda/D_\delta)$ where $D_\delta$ is
the cylinder length.
Combining with (\ref{cubevolume}) this gives
\begin{eqnarray}
\label{numberpointings}
\frac{V}{(1+z)^2 N_a N_d N_f} 
\;=\; \frac{d_T^2 d_H \Delta\Omega}{t_{tot}\nu_0}\;(dishes) \nonumber
\\
  =\; \frac{d_T^2 d_H \Delta\Omega}{t_{tot}\nu_0}
\,\frac{\lambda/D_\delta}{\Delta\delta}
\;(cylinders)
\end{eqnarray}

The values of $P_0$ thus differ for cylinders and dishes. 
For cylinders we have 
\begin{eqnarray}
\label{pnoisecyl}
P_0\;=\;
2.4\times10^4 (h^{-1}Mpc)^3
\left(\frac{T_{sys}}{50K}\right)^2
\left(\frac{d_T}{c/H_0}\right)^2
\left(\frac{\lambda/D_\delta}{0.525/100}\right)
\nonumber
\\ \times
\left(\frac{\Delta\alpha/t_{tot}}{2\pi/3\times10^7sec}\right)
%\end{displaymath}
%\begin{equation}\hspace*{30mm}\times
\frac
{\sqrt{\Omega_\Lambda + \Omega_M(1+z)^3}}
{(1+z)^2\,f_{HI}(z)^2}
%\left\langle |F(\vec{k},\vec{u})|^{-2} \right\rangle
\left(\frac{\Omega_{HI}h_{70}}{3.5\times10^{-4}}\right)^{-2}
\end{eqnarray}
For dishes we have a much larger $P_0$ because of the smaller sky coverage:
\begin{eqnarray}
\label{pnoisedish}
P_0\;=\;
4.6\times10^6 (h^{-1}Mpc)^3
\left(\frac{T_{sys}}{50K}\right)^2
\left(\frac{d_T}{c/H_0}\right)^2
%\left(\frac{\lambda/D_\delta}{0.525/100}\right)  \hspace*{10mm}
\hspace*{20mm}
\nonumber
\\ 
%\hspace*{10mm}
\times\left(\frac{\Delta\Omega/t_{tot}}{2\pi/3\times10^7sec}\right)
\frac
{\sqrt{\Omega_\Lambda + \Omega_M(1+z)^3}}
{(1+z)^2\,f_{HI}(z)^2}
%\left\langle |F(\vec{k},\vec{u})|^{-2} \right\rangle
\left(\frac{\Omega_{HI}h_{70}}{3.5\times10^{-4}}\right)^{-2}
%\nonumber
\end{eqnarray}

Using (\ref{meanetaformula}),
the expectation value of  the noise for the mode $\vec{k}$ is
\begin{equation}
\langle |\Gamma_{noi}(\vec{k})|^2 \rangle\;=\;
\frac{P_0 N_{\vec{u}}(\vec{k})}
{\left( \sum_{\vec{u}} |F(\vec{k},\vec{u})|
\right)^2}
=\frac{P_0}
{N_{\vec{u}}(\vec{k}) \langle |F(\vec{k},\vec{u})|\rangle_{\vec{u}}^2}
\label{gammanoise}
\end{equation}
where $P_0$ is given by (\ref{pnoisecyl}) or (\ref{pnoisedish})
and  $N_u(\vec{k})$ is the number of visibilities available for
estimating $\Gamma(\vec{k})$.
The noise is thus inversely proportional to 
$N_u(\vec{k})\langle F \rangle^2$.
We will see that the large value of $P_0$ for dishes compared to cylinders
will be compensated for by the larger value of $N_u$.

\subsection{The form factor}
\label{ffsec}

To estimate the form factor, we use the lobe function
\begin{displaymath}
L(\alpha^\prime,\delta^\prime,\nu^\prime) \;=\; 
\left(\frac{\sin \pi\alpha^\prime D_\alpha/\lambda}{\alpha^\prime}\right)^2\,
\left(\frac{\sin \pi\delta^\prime D_\delta/\lambda}{\delta^\prime}\right)^2\,
\left(\frac{\sin \pi\nu^\prime t_{int}}{\nu^\prime}\right)^2
\end{displaymath}
For a cylinder, $D_\alpha$ is the width, $D_\delta$ is the
length, and $t_{int}$ is the integrating time.
This lobe function is a reasonable approximation both for
the diffraction limited beam in right-ascension and for the FFT formed
beams in declination and frequency (if high frequency components are filtered
out).
The form factor for this lobe
is a product of triangle functions
$\Lambda(x)$  ($\Lambda(0)=1$,
$\Lambda(x)=0$ for $|x|>1$).
For $t_{int}>30\mu sec$, the frequency lobe is sufficiently narrow that
its Fourier transform is near unity for interesting values of $k$.
In this case, the form factor is the product of the two angular factors:
\begin{equation}
F(\vec{k},\vec{u}) \;=\; 
\Lambda\left(\frac{d_Tk_\alpha/2\pi-u_\alpha}{D_\alpha/\lambda}\right)\,
\Lambda\left(\frac{d_Tk_\delta/2\pi-u_\delta}{D_\delta/\lambda}\right)\,
\end{equation}
For cylinder arrays, $u_\delta=0$.
The most important characteristic of this form factor is that it 
vanishes as $k\rightarrow0$  (i.e. $k\ll 2\pi D_\alpha/\lambda d_T$).
This is because only correlations between separated pairs of cylinders
are used (no self-correlations) which makes the technique 
insensitive to noise drifts but at the price
of insensitivity to structure on high angular scales. 
Because of the triangular form of the form factor,
for the unpacked cylinder array, there are generally two values
of $u_\alpha$ that have non-vanishing $F(\vec{k},\vec{u})$ for a given 
$\vec{k}$:  the ones with $u_\alpha$ just above and just below
$k_\alpha/[2\pi D_\alpha/(\lambda d_T)]$.
For the other two arrays, there are more than two combinations.

\begin{figure}
\includegraphics[width=8.5cm]{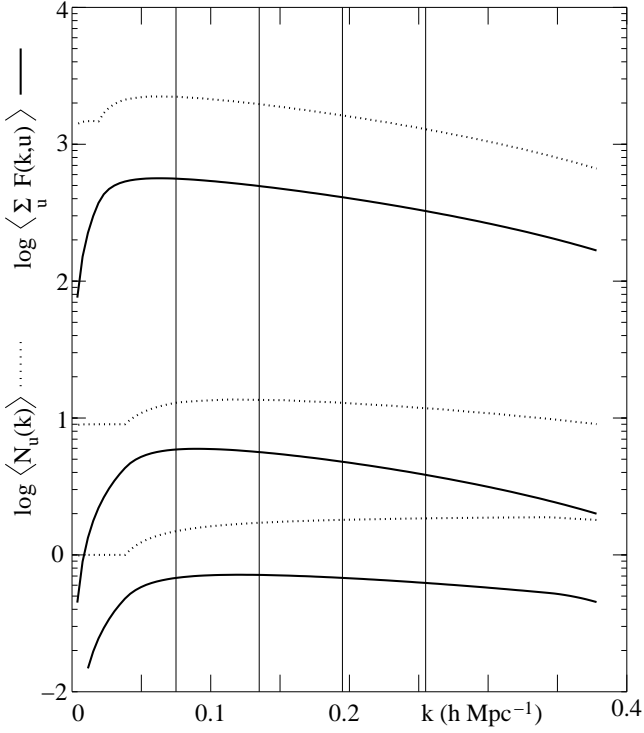}
\caption{ 
$\sum_{\vec{u}}F(\vec{k},\vec{u})$ (solid lines) 
and $N_{\vec{u}}(\vec{k})$ (dotted lines)
both averaged over the orientation of  $\vec{k}$.
The top pair of curves is for the dish array 
(configuration $c$ in Figure \ref{configfig}),
the middle pair for the packed cylinder array 
(configuration $b$ in Figure \ref{configfig}),
the bottom pair  for the unpacked cylinder array 
(configuration $a$ in Figure \ref{configfig}).
The vertical lines give the positions of the first four
BAO peaks.
}
\label{formfactorfig}
\end{figure}

Figure \ref{formfactorfig} plots  $N_u(\vec{k})$ and 
$\sum_{\vec{u}}F(\vec{k},\vec{u})$  
averaged over orientation of $\vec{k}$ for the three telescope 
configurations.
The dish array has many more pairs for a given $\vec{k}$ than the
cylinder array and this works to compensate for its higher noise.

\subsection{The power spectrum}
\label{noisepowersec}

The power spectrum $P(k)=P_{LSS}+P_{noi}$ 
can be found by calculating a
weighted average of the $|\Gamma(\vec{k})|^2$ for $k$ in
an interval $\Delta k$:
\begin{equation}
P(k) \;=\;  \frac
{\sum_{\vec{k}} W(\vec{k})|\Gamma(\vec{k})|^2}
{\sum_{\vec{k}} W(\vec{k})}
\end{equation}
where $\Gamma(\vec{k})$ is given by (\ref{meanetaformula}).
A reasonable choice is $W(\vec{k})=(\sum_{\vec{u}} F(\vec{k},\vec{u}))^2$
in which case
\begin{equation}
P(k) \;=\; \frac
{\sum_{\vec{k}}\,| \sum_{\vec{u}} 
F(\vec{k},\vec{u}) \Gamma(\vec{k},\vec{u})       |^2}
{\sum_{\vec{k}} \, (\sum_{\vec{u}} F(\vec{k},\vec{u}))^2  }
\end{equation}
Using (\ref{gammanoise}) 
we find  a simple expression for the noise:
\begin{equation}
P_{noi}(k) \;=\; \frac{P_0\left\langle N_u(\vec{k})\right\rangle} 
{\left\langle 
(\sum_{\vec{u}} F(\vec{k},\vec{u}))^2
\right \rangle  }
\label{noisedominated}
\end{equation}
where the averages are over $\vec{k}$.
As expected from (\ref{gammanoise}),
$P_{noi}(k)$ is  inversely proportional to 
$N_u \langle F \rangle^2$.

\begin{figure}
\includegraphics[width=8.5cm]{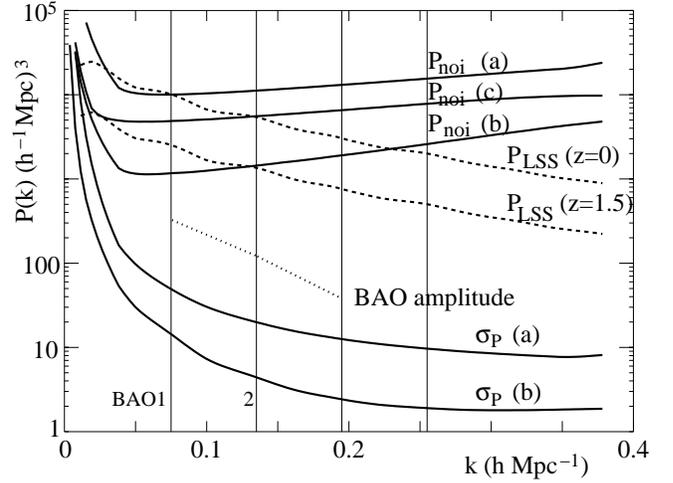}
\caption{The $P(k)$ and uncertainties for a four-month observation
of $2\pi sr$ at $z=1.5$ ($\Delta z=0.2$).
The  solid lines labeled $P_{noi}$
give the noise power
calculated with (\ref{noisedominated}) for the three configurations 
of Figure \ref{configfig}.
The  dashed lines shows the SDSS $P(k)$ at $z\sim0$ and
extrapolated to  $z=1.5$.
The dotted line shows the peak-to-peak amplitude
of the BAOs at $z=1.5$.
The solid lines labeled $\sigma_P$ shows the uncertainty in the
measured power spectrum for $\Delta k/k=0.2$
calculated with (\ref{sigma})
for configuration $a$ and $b$.
The vertical lines show the positions of the first
four BAO peaks.
}
\label{powerfig}
\end{figure}

Figure \ref{powerfig} shows the $P_{noi}(k)$
calculated in the three configurations for a four-month
observation of $2\pi sr$ at $z=1.5$  ($\Delta z=0.2$).
The noise for the packed array is considerably smaller than
that for the unpacked array because of the higher value
of $N_u$.  
The larger value of $P_0$ for the dishes is compensated by the
larger value of $N_u$ so the noise is comparable to that for
the cylinder arrays.  Note that the dishes 
accomplish this with fewer electronics channels, 400,
than  the 2500 (5000) channels needed for the unpacked (packed)
cylinders.

\begin{figure}
\includegraphics[width=8.5cm]{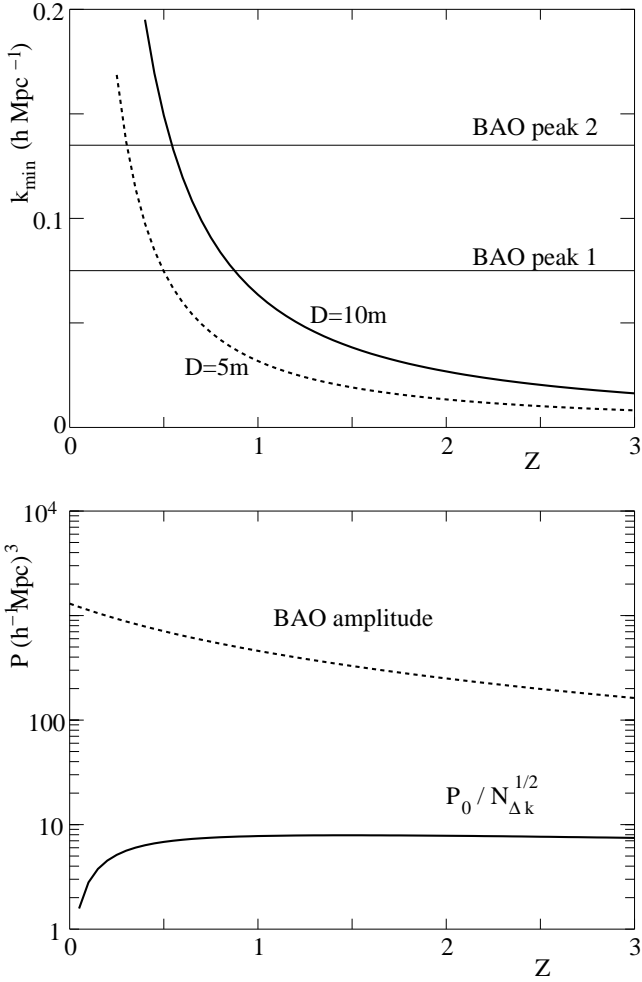}
\caption{The top panel shows
the $z$ dependence of $k_{min}=2\pi D/(\lambda d_T)$ 
for $D=10m$ (solid line)
and $D=5m$ (dashed line). 
The two horizontal lines show the position of the first two BAO peaks.
In the bottom panel, the solid line shows $P_0/\sqrt{N_{\Delta k}}$
for an exposure of $2\pi yr-sr$ cylinders in configuration $a$.  
The dashed line
shows the BAO peak-to-peak amplitude at the first peak
(lower panel).
}
\label{zdepfig}
\end{figure}

For all configurations,
the noise diverges as $k\rightarrow0$ because
of the filtering by the lobe, resulting in decrease in the form factor
and an increase in $P_{noi}(k)$.
The minimum $k$ with full sensitivity is
\begin{equation}
k_{min}=\frac{2\pi D_{min}/\lambda}{d_T(z)}
\;=\;
0.038 h\,Mpc^{-1}\,\frac{D_{min}/\lambda}{10m/0.525m}\frac{3.1Gpc}{d_T}
\end{equation}
where $D_{min}$ is
the minimum cylinder or
dish separation.  
For the numerical values of $d_T$ and $\lambda$
in this formula we have taken $z=1.5$ 
($\Omega_M,\Omega_\Lambda=0.27,0.73$) and the $D_{min}$ value
for configurations b and c of Fig. \ref{configfig}.
Figure \ref{zdepfig} plots $k_{min}(z)$ for $D_{min}=10m$
and $D_{min}=5m$.
To have good sensitivity at the first BAO peak,
$10m$ elements are acceptable at $z=1.5$ but smaller elements
are needed at $z=0.5$.

The uncertainty, $\sigma_P$, in the measured power spectrum averaged
over an interval $\Delta k$ is
\begin{equation}
\sigma_P \;=\; \frac
{\left\langle W(\vec{k})^2|\Gamma(\vec{k})|^4 \right\rangle^{1/2} }
{ \sqrt{N_{\Delta k}}\left\langle W(\vec{k}) \right\rangle} 
\label{sigma}
\end{equation}
where $N_{\Delta k}$ is the
number of modes in range $\Delta k$ given by (\ref{nmodes}).
If the noise dominates the power spectrum, this is
\begin{equation}
\sigma_P \;=\; \frac
{P_0 \left\langle N_u(\vec{k})^2 \right\rangle^{1/2} }
{ \sqrt{N_{\Delta k}}
\left\langle 
(\sum_{\vec{u}} F(\vec{k},\vec{u}))^2
\right\rangle} 
\label{noisedominatedsigma}
\end{equation}

Figure \ref{powerfig} shows 
%the noise and LSS power spectra and 
the resolution $\sigma_P$ calculated for a four-month
observation of $2\pi sr$ at $z=1.5$  ($\Delta z=0.2$) for the two 
cylinder configurations $a$ and $b$, the dish configuration giving
intermediate results.
The resolution is less than the BAO amplitude shown in the figure.
For the packed array $b$, $P_{noi}<P_{LSS}$ at the first
BAO peak so no improvement can be made by increasing the
observing time.  
For the unpacked array $c$, increasing the observing time
by a factor 10 would leak to a resolution similar to that
for the packed array.

If $P_{noi}>P_{LSS}$,
$\sigma_P$ is 
of order $(P_0/N_u(\vec{k}))/\sqrt{N_{\Delta k}}$.
The value of this quantity for unpacked cylinders 
and calculated at the first BAO peak 
is plotted as a function of redshift in Fig. \ref{zdepfig}.
Also shown is the peak-to-peak amplitude of the BAO oscillation.
The resolution relative to the BAO amplitude rises slowly with
redshift.

\section{Radio sources}
\label{radsourcesec}

Extragalactic radio sources and Milky Way Synchrotron radiation have
power law frequency spectra that allow them to be subtracted, 
in principle, from the HI spectrum.
We consider here radiation for extragalactic sources because
their random positions on the sky lead to a flux highly dependent
on angular position.
The radio source contribution to the sky brightness (\ref{skytemp})
is
\begin{displaymath}
T_{rs}(\alpha,\delta,\nu)\;=
\frac{c^2}{2k_b \nu^2}\,\sum_i \Phi_i 
\left(\frac{\nu}{\nu_0}\right)^{-\gamma_i}
\delta(\alpha-\alpha_i)\delta(\delta-\delta_i)
\end{displaymath}
where the sum is over radio sources $i$ of angular position 
$(\alpha_i,\delta_i)$, $1.42GHz$ flux (in $Jy$) 
$\Phi_i$, and spectral index is $\gamma_i$. 
The radio source power spectrum is determined by the Fourier transform:
\begin{eqnarray}
\Gamma_{rs}(\vec{k})\;=\;\frac{d_T^2 d_H(1+z)^2}{\bar{T}_{HI}\sqrt{V}\nu_0}\,
\hspace*{50mm}
\\ \hspace*{10mm} \times
\int d\alpha d\delta d\nu T_{rs}(\alpha,\delta,\nu)
e^{-i(d_T k_\alpha \alpha + d_T k_\delta \delta + d_H(1+z)^2 k_z \nu/\nu_0)}
\nonumber
\end{eqnarray}
The $\nu$ 
integral can be done by developing $\nu^{-\gamma}$ to first order around
the mean frequency in the box,
$\nu=\nu_0/(1+z)$.  The result is
\begin{eqnarray}
\label{gammars}
\Gamma_{rs}(\vec{k}) \;=\;-i(-1)^n \exp(-id_zk_z/(1+z))
\hspace*{20mm}
 \\ 
%\hspace*{10mm}
\frac{d_T^2 }{k_z \sqrt{V}}\,
\frac{\Delta\nu}{\nu_0}\,
(\gamma+2)(1+z)^{\gamma+3}
\frac{c^2}{2k_b \nu_0^2 \bar{T}_{HI}} \,
\eta(k_\alpha, k_\delta)
\nonumber
\end{eqnarray}
where $n=n_z$ is defined by (\ref{skytemp}),
$\gamma\sim0.7$ is the mean spectral index \citep{binneymerrifield}, 
and $\eta$ is a sum over the radio sources in the angular
region $\Delta\alpha\Delta\delta$:
\begin{equation}
\eta(k_\alpha, k_\delta)\;=
(\gamma+2)^{-1}\,\sum_i (\gamma_i+2) (1+z)^{\gamma_i-\gamma}\Phi_i
e^{-id_T(k_\alpha \alpha + k_\delta \delta )}
\end{equation}
(Note: $z$ is the redshift of the box, not of the radio source).
The power in the mode $\vec{k}$ is then
\begin{eqnarray}
|\Gamma_{rs}(\vec{k})|^2 \;=\; 
\frac{d_T^2}{k_z^2 d_H}\,
\Delta z\,
%\frac{\Delta\nu}{\nu_0}\,
(\gamma+2)^2\,
\hspace*{30mm}
\nonumber
\\ 
\hspace*{15mm}
(1+z)^{2(\gamma+1)} 
%(1+z)^{2(\gamma+3)} 
\left(\frac{c^2}{2k_b \nu_0^2 \bar{T}_{HI}}\right)^2\,
%\\
\Delta\Omega^{-1}\,
|\eta(k_\alpha,k_\delta)|^2
\end{eqnarray}
This spectrum is highly anisotropic because fluxes are correlated
in frequency but not in angular direction.

The expectation value of $|\eta|^2$ is
\begin{equation}
\Delta\Omega^{-1}\,
\left\langle |\eta(k_\alpha,k_\delta)|^2 \right\rangle \;=\;
\int_0^{\infty} d\Phi \frac{dN}{d\Phi d\Omega} \Phi^2
\;\sim 67.7\, Jy^2\,sr^{-1}
\end{equation}
where we use the spectrum from \citet{jackson}.
Half of the integral
comes from the $\sim5000$ sources per steradian with
fluxes greater than 0.06Jy.
This gives
\begin{equation}
\frac{c^2}{2k_b \nu_0^2 }\,
\left(\Delta\Omega^{-1}\,
\left\langle |\eta(k_\alpha,k_\delta)|^2 \right\rangle
\right)^{1/2} \;\sim\;
0.000133\,K
\end{equation}
For $z=1.5$,  $\gamma=0.7$, $\Delta z=0.2$ we then have
\begin{equation}
\left\langle
|\Gamma_{rs}(\vec{k})|^2 
\right\rangle \;=\;
7.0\times10^6 (h^{-1}Mpc)^3 \left( 
\frac{0.075h\,Mpc^{-1}}{k_z}\right)^2
\end{equation}
where the average is over $(k_\alpha,k_\delta)$ at fixed
$k_z$.
Since this is considerably larger than $P_{noi}$ and $P_{LSS}$,
$\Gamma_{rs}(\vec{k})$ must be subtracted 
from the measured $\Gamma(\vec{k})$ for each $\vec{k}$.
This can be done by fitting the observed  $\Gamma(\vec{k})$ with
the $f(k_\alpha,k_\delta)/k_z$  form given by (\ref{gammars}).
In principle, this subtraction can be done with a precision 
$\sigma_\Gamma\sim P_{noi}/\sqrt{N_z}$ where $N_z$ is the number of radial
modes over which the determination of  $\Gamma_{rs}$ is made.
If this precision can be reached in practice, the subtracted power spectrum
is not degraded below the spectrum in the absence of radio sources.

\section{Sensitivity to cosmological parameters}
\label{cosmosec}

In Sec. \ref{recsec}, 
we studied the impact of the various telescope configurations 
on power spectrum reconstruction.  Fig.~\ref{powerfig} shows the various power 
spectra, and it allows us to rank visually the configurations in terms of 
electronics noise fraction. 
The differences in $P_{noi}$ will translate into differing precisions
in the reconstruction of the BAO peak positions and in
the estimation of cosmological parameters.

\subsection{BAO peak precision}

In order to estimate the precision with which BAO peak positions can be
measured, we have generated power spectra that are the sum of the expected
HI power spectra and the noise spectra calculated in the 
Section \ref{noisesec}.
The peaks in the generated spectra were then determined by a 
fitting procedure
and the reconstructed peak positions compared with the 
generated peak positions.

To this end,
we  used a method similar to the one established in \citet{Blake}.
We  generated power spectra for
 slices of Universe with a half-sky coverage and a redshift depth,
 $\Delta z=0.2$ for  $0<z<1.6$. 
The power spectrum used in the simulation is  
the sum of HI signal term, corresponding to (\ref{eq:signal}) 
and noise term 
derived from (\ref{gammanoise}).  The simulated power spectra HI is:
%\begin{equation}
\begin{eqnarray}
\label{eq:signal}
\frac{P_{HI}(\kperp,\kpar)}{P_{ref}(\kperp,\kpar)} = 
1\; + 
\hspace*{40mm}
\nonumber
\\ \hspace*{20mm}
A\, k \exp \bigl( -(k/\tau)^\alpha\bigr)
\sin\left( 2\pi\sqrt{\frac{\kperp^2}{\koperp^2} + 
\frac{\kpar^2}{\kopar^2}}\;\right)
\end{eqnarray}
%\end{equation}
where $k=\sqrt{\kperp^2 + \kpar^2}$, the parameters $A$, $\alpha$ and $\tau$ 
are adjusted to the  formula presented in 
\citet{Eisens1998}. $P_{ref}(\kperp,\kpar)$ is the 
envelop curve of the HI power spectrum without baryonic oscillations. 
The parameters $\koperp$ and $\kopar$ 
are the inverses of the oscillation periods in k-space. 
In the simulations, we used the following values for  these 
parameters: $A=1.0$, $\tau=0.1\hMpcm$, 
$\alpha=1.4$ and $\koperp=\kopar=0.060\hMpcm$.

Each simulation is performed for a given set of parameters 
which are: the system temperature,$T_{sys}$, an observation time, 
$T_{obs}$, an average redshift and a redshift depth, $\Delta z=0.2$. 
Then,  each simulated  power spectrum  is fitted with a two dimensional 
normalized function $P_{tot}(\kperp,\kpar)/P_{ref}(\kperp,\kpar)$ which is 
the sum of the signal power spectrum of (\ref{eq:signal}) and the normalized 
noise power spectrum defined by (\ref{gammanoise}) multiplied by a 
linear term,  $a_0+a_1k$. The upper limit $k_{max}$ in $k$ of the fit 
corresponds to the approximate position of the linear/non-linear transition. 
This limit is established on the basis of the criterion discussed in  
\citet{Blake}. 
In practice, we used for the redshifts 
$z=0.5,\,\, 1.1$ and  1.5 respectively $k_{max}= 0.145 \hMpcm,\,\, 0.19\hMpcm$ 
and $0.23 \hMpcm$. 
 
\begin{figure}[htbp]
\begin{center}
\includegraphics[width=8.5cm]{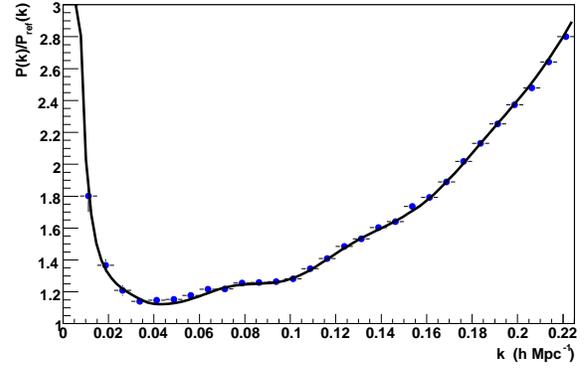}
\caption{1D projection of the power spectrum for one simulation. 
The HI power spectrum is divided by an envelop curve $P(k)_{ref}$ 
corresponding to the power spectrum without baryonic oscillations. 
The dots represents one simulation for a "packed" array of cylinders  
with a system temperature,$T_{sys}=50$K, an observation time, 
$T_{obs}=4 {\rm month}$, 
a solid angle of $2\pi sr$,
an average redshift, $z=1.5$ and a redshift depth, $\Delta z=0.2$. 
The solid line is the result of the fit to the data.}
\label{fig:fitOscill}
\end{center}
\end{figure}

\begin{figure}[htbp]
\begin{center}
\includegraphics[width=9.0cm]{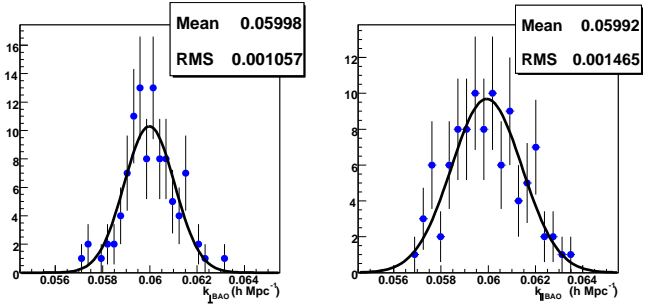}
\caption{ Distributions of the reconstructed 
wavelength  $\koperp$ and $\kopar$ 
respectively, perpendicular and parallel to the line of sight
for simulations as in Fig. \ref{fig:fitOscill}. 
The fit by a Gaussian of the distribution (solid line) gives the 
width of the distribution  which represents the statistical error 
expected on these parameters.}
\label{fig:McV2}
\end{center}
\end{figure}

Figure \ref{fig:fitOscill} shows the result of the fit for 
one of theses simulations. 
Figure \ref{fig:McV2} histograms the recovered values of  $\koperp$ and $\kopar$
for 100 simulations.
The widths of the two distributions give an estimate 
the statistical errors.

\begin{figure}[htbp]
\begin{center}
\includegraphics[width=8.5cm]{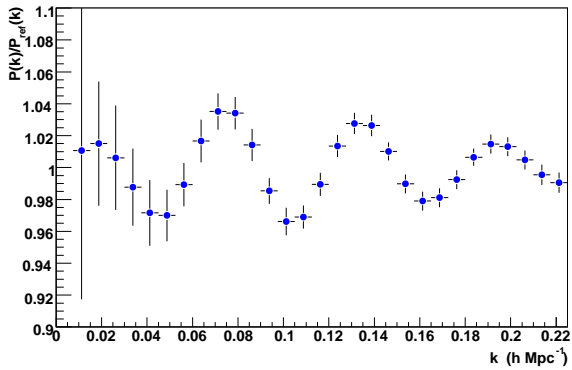}
\caption{1D projection of the power spectrum averaged over 100 simulations
of the packed cylinder array $b$. 
The simulations are performed for the following conditions: a system 
temperature, $T_{sys}=50$K, an observation time, $T_{obs}=1 {\rm year}$, 
a solid angle of $2\pi sr$,
an average redshift, $z=1.5$ and a redshift depth, $\Delta z=0.2$. 
The HI power spectrum is divided by an envelop curve $P(k)_{ref}$ 
corresponding to the power spectrum without baryonic oscillations 
and the background estimated by a fit is subtracted. The errors are 
the RMS of the 100 distributions for each $k$ bin and the dots are 
the mean of the distribution for each $k$ bin. }
\label{fig:AverPk}
\end{center}
\end{figure}

In addition, in the fitting procedure, both the parameters modeling the 
signal $A$, $\tau$, $\alpha$ and the parameter correcting the noise power 
spectrum $(a_0,a_1)$ are floated to take into account the possible 
ignorance  of the signal shape and the uncertainties in the 
computation of the noise power spectrum. 
In this way, we can correct possible imperfections and the 
systematic uncertainties are directly propagated to statistical errors 
on the relevant parameters  $\koperp$ and $\kopar$. By subtracting the 
fitted noise contribution to each simulation, the baryonic oscillations 
are clearly observed, for instance, on Fig.~\ref{fig:AverPk}.

%\subsection{Results}

In our comparison of the various telescope configurations, we have considered 
the following cases for $\Delta z=0.2$ slices with $0<z<1.6$.
\begin{itemize}
\item {\it Simulation without electronics noise}: the errors of power 
spectrum are directly related to the number of modes in the $\Delta z=0.2$ 
slice of 2$\pi$sr (\ref{nmodes}). 
\item {\it Unpacked cylinder array} (configuration a)
\item {\it Packed cylinder array}: (configuration b). 
\item {\it Dish array}: (configuration c). 
\end{itemize}

Table \ref{tab:ErrorOnK} summarizes the result. The ranking 
of the three configurations is the same as that 
which we can deduce from the noise levels in  
Fig.~\ref{powerfig}: the best sensitivity is obtained with the packed 
cylinder array, then with the packed dish array and finally with the 
unpacked cylinder array. However,   none of the three configurations 
is limited by the cosmic variance with a four-month observation. As 
the errors scales with  $1/T_{obs}$, configurations a) and c) could 
be significantly improved if the observation time is increased up to one year. 

\begin{table*}[ht]
\caption{Sensitivity on the measurement of $\koperp$ and $\kopar$ as a 
function of the redshift $z$ for various telescope configurations. 
$1^{\rm st}$ row: simulations without noise with pure cosmic variance; 
$2^{\rm nd}$ 
row: simulations with same electronics noise for a telescope in an 
unpacked cylinder array configuration a); 
$3^{\rm th}$ row: simulations 
with same electronics noise for a telescope in a packed cylinder array 
configuration b); 
$4^{\rm th}$ row: simulations with same electronics 
noise for a telescope in a packed dish array configuration c).}
\begin{center}
\begin{tabular}{lc|c c c c c c c c }
\multicolumn{2}{c|}{$\mathbf z$ }& \bf 0.1 & \bf 0.3 &  \bf 0.5 & \bf 0.7 & \bf 0.9 &  \bf 1.1 &  \bf 1.3 &  \bf 1.5\\
\hline\hline
\bf No Noise & $\sigma(\koperp)/\koperp$  (\%) & 16 & 4.5 & 2.3 & 1.6 & 1.1 & 0.90 & 0.79 &0.73 \\
 & $\sigma(\kopar)/\kopar$  (\%) & 18 & 5.8 & 3.1& 2.3 & 1.6 & 1.3 & 1.2 &1.1\\
 \hline
% \bf Flat Noise & $\sigma(\koperp)/\koperp$  (\%) & 16 & 5.2  & 3.4 & 3.3 &  3.1 &  3.2 & 3.3 & 3.5\\
% & $\sigma(\kopar)/\kopar$  (\%) & 18 & 6.5 & 5.1 & 5.0 & 5.0 & 5.1 & 5.2 & 5.5\\
% \hline
 \bf  a) Unpacked cylinder array   & $\sigma(\koperp)/\koperp$  (\%) & -& - & 19 & 16 &  12 &  9.8 & 9.3 & 9.8\\
 (4-months/redshift)& $\sigma(\kopar)/\kopar$  (\%) & - & - &  24 & 24 & 15 &  12 &  11 & 11\\
 \hline
 \bf  b) Packed cylinder array & $\sigma(\koperp)/\koperp$  (\%) & - &  9.0 &  3.2 & 2.1 & 1.6 &  1.6 & 1.7 & 1.9\\
 (4-months/redshift)& $\sigma(\kopar)/\kopar$  (\%) & - & 15 & 6.3 & 3.8 & 2.4 & 2.2 & 2.4 & 2.6\\
 \hline
 \bf  c) Packed Dish array & $\sigma(\koperp)/\koperp$  (\%) & - & 11  &  7.2 & 6.5 & 5.7 & 5.5 & 5.5  & 5.6 \\
 (4-months/redshift)& $\sigma(\kopar)/\kopar$  (\%) & - &  24 & 11 & 8.6 &  6.7 &  6.4 & 6.9 &  6.9\\
 \hline
 \bf  b) Packed cylinder array & $\sigma(\koperp)/\koperp$ (\%) & - &  - & 2.0 & 1.5 & 1.1 & 1.1 & 1.0 & 1.1 \\
 (4-year optimized)              & $\sigma(\kopar)/\kopar$ (\%) & - & - & 4.3 & 2.7 & 1.7 & 1.6 & 1.4 & 1.4 \\
\end{tabular}
\end{center}
\label{tab:ErrorOnK}
\end{table*}%

\subsection{Expected sensitivity on $w_0$  and $w_a$}

\begin{figure}
\begin{center}
\includegraphics[width=8.5cm]{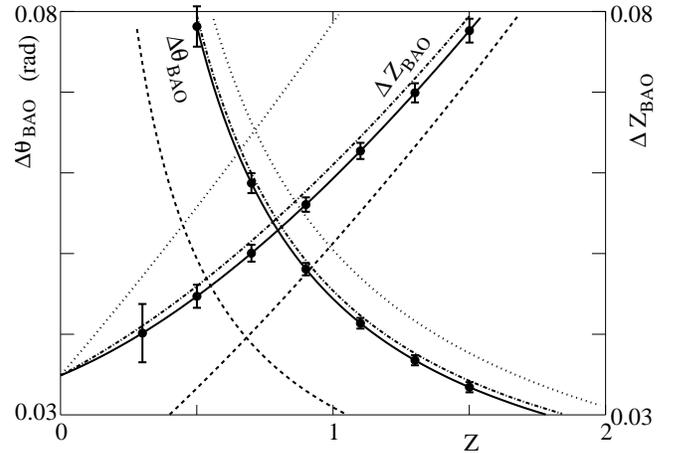}
\caption{
The two ``Hubble diagrams'' for BAO experiments.
The four falling curves give the angular size of the acoustic horizon
(left scale) and the four 
rising curves give the redshift interval of the horizon (right scale).
The solid lines are for 
$(\Omega_M,\Omega_\Lambda,w)=(0.27,0.73,-1)$,
the dashed  for 
$(1,0,-1)$
the dotted for 
$(0.27,0,-1)$, and
the dash-dotted  for 
$(0.27,0.73,-0.9)$,
The error bars on the solid curve correspond to the four-month run
(packed array)
of Table \ref{tab:ErrorOnK}.
 }
\label{fig:hubble}
\end{center}
\end{figure}

The observations give the HI power spectrum in
angle-angle-redshift space rather than in real space.
The inverse of the peak positions  in the observed power spectrum therefore  
gives the angular and redshift intervals corresponding to the
sonic horizon.
The peaks in the angular spectrum are proportional to $d_T(z)/s$
and those in the redshift spectrum to $d_H(z)/s$.
The quantities $d_T$, $d_H$ and $s$ all depend on
the cosmological parameters.
Figure \ref{fig:hubble} gives the angular and redshift intervals
as a function of redshift for four cosmological models.
The error bars on the lines for
$(\Omega_M,\Omega_\Lambda)=(0.27,0.73)$
correspond to the expected errors 
on the peak positions
taken from Table \ref{tab:ErrorOnK}
for the four-month runs with the packed array.
We see that with these uncertainties, the data would be able to 
measure $w$ at better than the 10\% level.

\begin{figure}[htbp]
\begin{center}
\includegraphics[width=8.5cm]{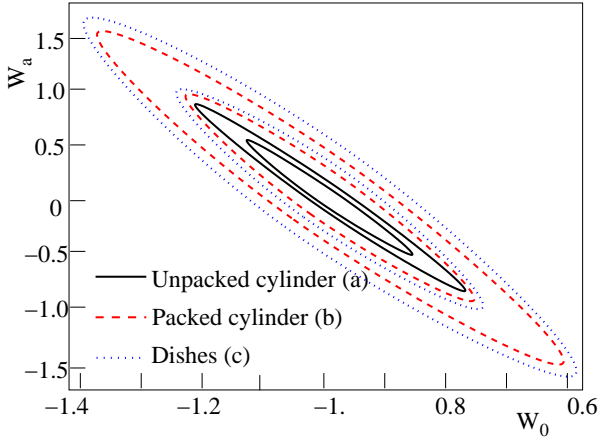}
\caption{$1\sigma$ and $2\sigma$ confidence level contours  in the parameter 
plane $(w_0,w_a)$ for three configurations : a) unpacked array of north-south 
oriented cylinders (black line), b)   packed array of north-south oriented 
cylinders (blue line)  and c) packed array of  dishes (red line).}
\label{fig:w0wa}
\end{center}
\end{figure}

To estimate the sensitivity 
to parameters describing dark energy equation of 
state, we follow the procedure explained in 
\citet{Blake}. We can introduce the equation of 
state of dark energy, $w(z)=w_0 + w_a\cdot z/(1+z)$ by 
replacing $\Omega_\Lambda$ in the definition of $d_T (z)$ and $d_H (z)$, 
(\ref{dTdH}) by:
\begin{equation}
\Omega_\Lambda = \Omega_{\Lambda}^0 \exp \left[ 3  \int_0^z   
\frac{1+w(z^\prime)}{1+z^\prime } dz^\prime  \right]
\end{equation}
where $\Omega_{\Lambda}^0$ is the present-day dark energy fraction with 
respect to the critical density. 
Using the relative errors on  $\koperp$ and  $\kopar$ given in 
Tab.~\ref{tab:ErrorOnK}, we can compute the Fisher matrix for  
five cosmological parameter: $(\Omega_m, \Omega_b, h, w_0, w_a)$. 
Then, the combination of this BAO Fisher 
matrix with the Fisher matrix obtained for Planck mission, allows us to 
compute the errors on dark energy parameters.
The Planck Fisher matrix is
obtained for the 8 parameters (assuming a flat universe): 
$\Omega_m$, $\Omega_b$, $h$, $w_0$, $w_a$,
$\sigma_8$, $n_s$ (spectral index of the primordial power spectrum) and
$\tau$  (optical depth to the last-scatter surface).
The expected errors and the 
Figure of Merit, the inverse of the area in the 95\% confidence level 
contours (see Fig.~\ref{fig:w0wa}) are summarized in 
Tab.~\ref{tab:ErrorOnw0wa}. The ranking between the various configurations 
in terms of performances is consistent with the level of electronics noise  
observed on the power spectra of Fig.~\ref{powerfig}.

\begin{figure}[htbp]
\begin{center}
\includegraphics[width=8.5cm]{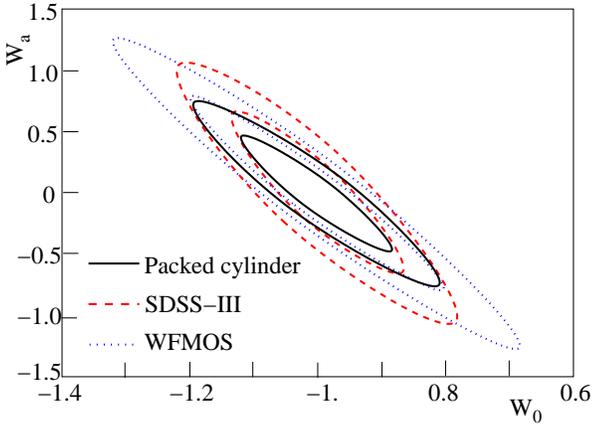}
\caption{$1\sigma$ and $2\sigma$ confidence level contours  in the 
parameter plane $(w_0,w_a)$ for three BAO projects:  WFMOS project 
(blue dotted line), SDSS-III (red dashed line)  and packed cylinder 
array with HI intensity mapping (black solid line).}
\label{fig:Compw0wa}
\end{center}
\end{figure}

For an optimized project over a redshift range, $0.4<z<1.6$, with a total 
observation time of 4 years, the packed cylinders have a 
precision of  6\% on $w_0$ and 25\% on $w_a$. Finally, Fig.~\ref{fig:Compw0wa} 
shows a comparison of different BAO projects, with a set of priors on 
$(\Omega_m, \Omega_b, h)$ corresponding to the expected precision on 
these parameters in early 2010's. This BAO project based on HI intensity 
mapping is clearly competitive with the next generation of optical 
surveys such as SDSS-III \citep{sdss3} or WFMOS \citep{wfmos}.

\begin{table}[htdp]
\caption{Sensitivity on $w_0$  and $w_a$ and Figure of Merit (FoM) for 
various telescope configurations. 
$1^{\rm st}$ row: simulations without 
noise with pure cosmic variance; 
$2^{\rm nd}$ row: simulations with same 
electronics noise for a telescope in an unpacked cylinder array 
configuration a); 
$3^{\rm th}$ row: simulations with same electronics 
noise for a telescope in a packed cylinder array configuration b); 
$4^{\rm th}$ row: simulations with same electronics noise for a telescope 
in a packed dish array configuration c); $6^{\rm th}$ row: Configuration b) 
for an optimized 4-year survey with $0.4<z<1.6$.}
\begin{center}
\begin{tabular}{l|c|c|c}
& $\sigma(w_0)$    &  $\sigma(w_a)$      & FoM (95\%) \\
\hline \hline
\bf No Noise & 0.059 & 0.24  & 136 \\
\bf a) Unpacked cylinder array  & 0.17 & 0.69 & 10.1 \\
\bf b) Packed cylinder array  & 0.091 & 0.36 & 69.1 \\
\bf c) Packed Dish array  & 0.15 & 0.64 & 13.6 \\
\bf Packed cylinder array - 4 years  & \bf 0.063 &  \bf 0.25 &  \bf 119 \\
\end{tabular}
\end{center}
\label{tab:ErrorOnw0wa}
\end{table}%

\section{Conclusion}

In this paper we have discussed the measurement of the
HI power spectrum with interferometric surveys.
By presenting a simplified procedure for reconstructing the
Fourier space map of the GHz sky, we derived expressions
for the noise power spectrum.  Adding this spectrum to the
expected HI power spectrum, we determined with what 
precision the positions of the BAO peaks can be measured.
This led to  a sensitivity to cosmological parameters
that is competitive with other BAO projects.

In calculating the noise power spectrum, we considered
three interferometer arrays using cylinders or dishes.
Packed-cylinder and packed-dish arrays have comparable
noises so the choice for one or the other must be based
on cost and technical considerations,
with cylinder arrays having simpler mechanical design but
requiring more receivers.

\begin{acknowledgements}
We thank J.B. Peterson and U.-L. Pen
for triggering our interest in cosmological HI interferometry
and for many interesting discussions.
\end{acknowledgements}

\bibliographystyle{aa}

\end{document}